# Predictive Model of Hydrogen Trapping and Bubbling in Nanovoids in BCC Metals

Jie Hou[1,2,3], Xiang-Shan Kong, Xuebang Wu[1], Jun Song[3], C. S. Liu

[1]Key Laboratory of Materials Physics, Institute of Solid State Physics, Chinese Academy of Sciences, P. O. Box 1129, Hefei 230031, P. R. China

[2]University of Science and Technology of China, Hefei 230026, P. R. China

[3]Department of Mining and Materials Engineering, McGill University, Montreal, Quebec H3A 0C5, Canada.

**Abstract:**

Interplay between hydrogen and nanovoids, despite long-recognized as a central aspect in hydrogen-induced damages in structural materials, remains poorly understood. Focusing on tungsten as a model BCC system, the present study, for the first time, explicitly demonstrated sequential adsorption of hydrogen adatoms on Wigner-Seitz squares of nanovoids with distinct energy levels. Interaction between hydrogen adatoms on the nanovoid surface is shown to be dominated by pairwise power law repulsion. A predictive model was established for quantitative prediction of configurations and energetics of hydrogen adatoms in nanovoids. This model, further combined with equation of states of hydrogen gas, enables prediction of hydrogen molecule formation in nanovoids. Multiscale simulations based on the predictive model were performed, showing excellent agreement with experiments. This work clarifies fundamental physics and provides full-scale predictive model for hydrogen trapping and bubbling in nanovoids, offering long-sought mechanistic insights crucial for understanding hydrogen-induced damages in structural materials.

Being the most abundant element in the known universe and a typical product from corrosion, hydrogen (H) exists virtually in all service environments. The exposure of metallic materials to H-rich environments can result in numerous structural damages, including H induced cracking[1,2], H induced surface blistering/flaking[3-7], and H induced porosity/swelling[8-13], among others. These damages undesirably degrade the structural and mechanical integrity of materials [14-16], often causing premature and even catastrophic failures[4,6,17], and thus jeopardizing safety and efficiency of many applications. It's generally believed that these damages originate from interactions between H and various lattice defects. One key issue among those interactions is the H interplay with nanovoids, which promotes the formation and growth of mesoscale H bubbles and consequently leads to experimentally observable failures in structural materials[3-11, 15, 16, 18].

The development of advanced micrographic techniques enables direct observation of H bubble structures at scales around tens of nanometers[4, 18]. However, atomic details of H bubble nucleation, growth and agglomeration processes are difficult or even impossible to observe *in-situ*. An alternative method to study H bubble formation is to measure H thermal desorption rates during isochronal annealing[19-23], which provides energetic information of H de-trapping from nanovoids in metals. Nevertheless, H thermal desorption spectra usually relate to multiple types of crystal defects at different depths[24], therefore difficult to interpret. Multiscale simulations provide a way to circumvent such limitation to reveal those missing details[25-27]. At the fundamental level, *ab initio* calculations based on density functional theory (DFT) have been widely used to study the atomistic behaviors of H in vacancies[28-42], showing that individual H atoms preferentially reside around tetrahedral interstitial sites or octahedral interstitial sites[28-34, 36, 37] on vacancy surfaces as adatoms, and further accumulation of H will quickly saturate the vacancy due to repulsion between H adatoms[28-38]. However, most of previous DFT studies focused only on H in monovacancies[28-35, 39, 40], rendering a knowledge gap for understanding H in general nanovoids. To fill this gap, notable efforts have been carried out using large-scale molecular dynamics (MD) simulations based on empirical interatomic potentials to investigate H trapping in nanovoids. Nonetheless, vastly different predictions of H trapping behaviors were obtained depending on the interatomic potentials used [43, 44]. Furthermore, many of the available interatomic potentials were designed for simulating bulk conditions and cannot reproduce $H_2$ molecule formation in bubbles that observed in experiments[45-47]. Consequently, it remains elusive which of those H trapping behaviors reported, if any, are physically meaningful and/or relevant to H bubble formation.

Recently, some DFT studies showed that pressurized $H_2$ molecules may form in the nanovoid core after a prerequisite stage of non-interacting H adatoms adsorption on nanovoid surfaces[48, 49], and proposed ways to characterize these two stages. Unfortunately, impeded by the structure complexity of nanovoid surfaces, the physical rules governing H trapping in nanovoids remain not understood, and more importantly the H-H interaction was overlooked by those previous studies. This absence of physical models and predictability prevents accurate analysis of H behaviors in nanovoids, rendering a formidable challenge to obtain mechanistic and multiscale insights toward H bubble formation in metals. In this study, we clarify the behavior of H in nanovoids in body-centered cubic (BCC) metals. We choose tungsten (W) as a representative for its important roles in withstanding H plasma in fusion reactors, with benchmark calculations for other typical BCC systems, e.g., Mo, Cr and alpha-Fe, also performed to confirm the generality of our conclusions. We explicitly demonstrated the sequential adsorption of H adatoms on square Wigner-Seitz surfaces of nanovoids with distinct energy levels, and proposed a power law to describe the interaction among H adatoms. A predictive model was then established for determining energetics and stable configurations of multiple H adatoms in a nanovoid. Further combined with the equation of state of pressurized $H_2$, this model enables quantitative assessment of the competition between H adatoms and $H_2$ molecules, and prediction of $H_2$ molecule formation in nanovoids. Multiscale simulations based on the model show excellent agreement with previous deuterium (*D*) thermal desorption experiments. The present study clarifies fundamental physical rules and provides full-scale prediction for H trapping and bubbling in nanovoids in BCC metals.

**Results**

**Structures of H adatoms trapped in nanovoids.** The general patterns of H trapping in nanovoids have been examined by comprehensive *ab initio* molecular dynamics (MD) simulations. H atoms were sequentially introduced into a nanovoid ($V_m$) to form a H-nanovoid clusters ($V_m H_n$), where $m = 1\sim8$ and $n \geq 0$ respectively denote the number of vacancies constituting the nanovoid and the number of H atoms enclosed therein. Similar to previous studies [48, 49], we observed H adatom adsorption on surface of all nanovoids and $H_2$ molecule formation in the core of large nanovoids ($V_m H_n$ with $m \geq 3$), evidenced by two distinct pairing states with H-H separation distances at around 1.94 Å and 0.75 Å, respectively (see Supplementary Section S1).

Fig. 1a illustrates the spatial locations of the high H probability density (i.e., the probability of finding a H in unit volume) regions in nanovoids during H adding. Our results indicate that H adatoms prefer to stay on the square surfaces of Wigner-Seitz cells of nanovoids (below those square units are referred as Wigner-Seitz squares), particularly near their vertexes (i.e., the tetrahedral interstitial sites). As shown in Fig. 1a, each Wigner-Seitz square is enclosed by six metal sites. While other metal sites are beyond 2.9 Å away from H adatoms on the Wigner-Seitz square, significantly larger than typical transition metal-H bond length $(1.5-2$ Å$)^{50}$. Therefore, it is reasonable to regard that energetic behaviors of H on a Wigner-Seitz square are mainly influenced by these six metal sites enclosing the square[29]. Note for a Wigner-Seitz square on nanovoid surfaces, part of these six metal sites are occupied by vacancies. Consequently, the complex nanovoid surface can be categorized into 5 different Wigner-Seitz squares, with $i$ and $j$ denoting vacancy numbers on two types of neighboring metal sites (see Fig. 1a). Fig. 1b presents an example of the average number of H atoms on each type of these Wigner-Seitz squares obtained from the *ab initio* MD simulations. Despite certain temperature-induced fluctuation (at relatively high temperature of 600K), it shows that H adatoms sequentially occupy different types of Wigner-Seitz squares (e.g., in the order of *ij*=22,12, 11, 10 in Fig. 1b) until all Wigner-Seitz squares are filled. After a certain prerequisite surface trapping, H starts to form $H_2$ molecules in the core of large nanovoids ($V_{\geq 3}$) (see Supplementary Section S1). This spatial preference clearly indicates an energy difference among H on different Wigner-Seitz squares and $H_2$ in the core.

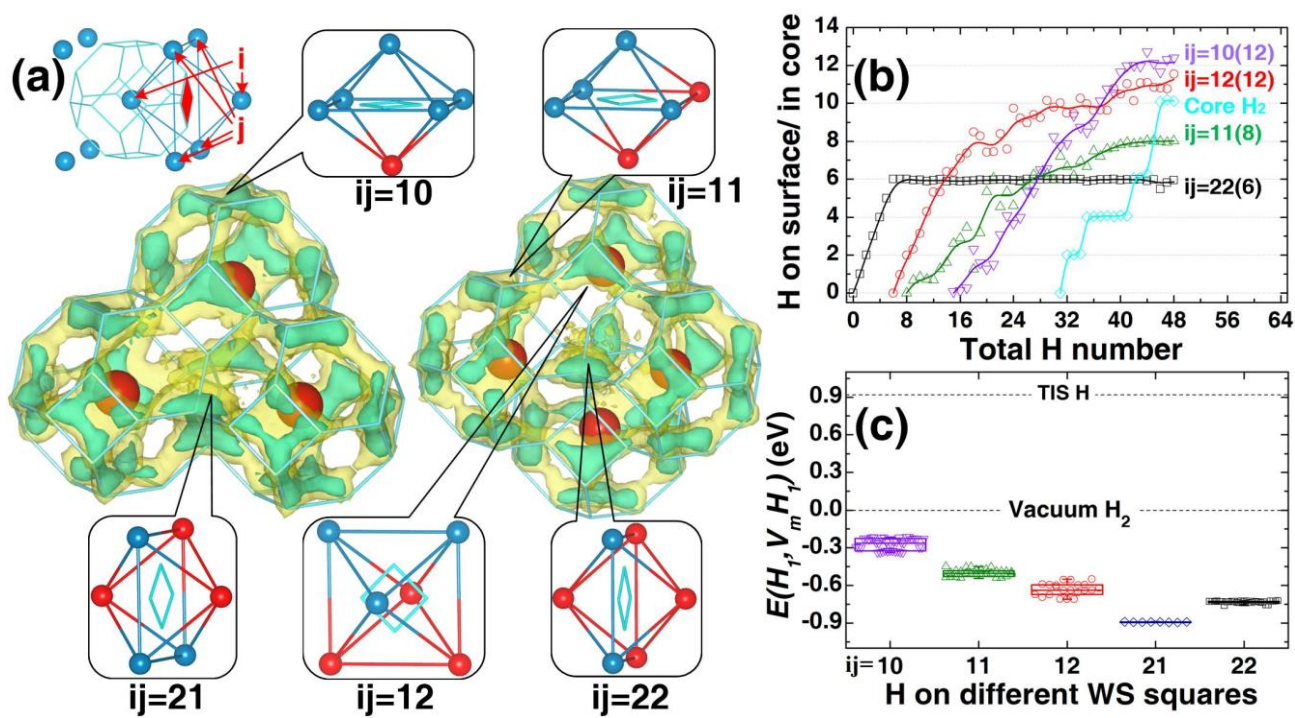

Figure. 1 (a) H probability density isosurfaces in $V_3$ and $V_4$ nanovoids during H adding under 600K. Blue and red spheres denote metal and vacancy sites respectively, with bright green lines indicating edges of Wigner-Seitz cells, *i* and *j* denotes vacancy numbers on two types of neighboring metal sites around a Wigner-Seitz square. W atoms (not shown) are fixed at their ideal BCC sites. (b) Average H numbers on different Wigner-Seitz squares (as adatoms) or in core (as molecules) in $V_8$ under 600K, shown in parenthesis are numbers of corresponding Wigner-Seitz squares on $V_8$ surface (note $V_8$ does not contain ij=21 Wigner-Seitz squares, see Supplementary Table SI), lines are just to guide the eye. (c) Trapping energy of a single H adatom (i.e., $E(H_1, V_m H_1)$, see Eq. 1) on all symmetrically irreducible square Wigner-Seitz squares in $V_1$-$V_8$. The lines marked as "TIS H" and "Vacuum $H_2$" represent a H atom at the tetrahedral interstitial site in bulk metal lattice and in a $H_2$ molecule in vacuum, respectively. Additional results for other nanovoids can be found in Supplementary Section S1.

To quantitatively analyze the energetics of H, we define the trapping energy of $k$ number of H in

a $V_mH_n$ cluster as:

$$E(H_k, V_mH_n) = E^{tot}(V_mH_n) - E^{tot}(V_mH_{n-k}) - \frac{k}{2}E^{tot}(H_2), \qquad (1)$$

where $E^{tot}(V_mH_{n-k})$ and $E^{tot}(V_mH_n)$ are the total energies of the reference metal matrix containing the stable $V_mH_{n-k}$ and $V_mH_n$ clusters, respectively before and after the introduction of the $k$ H atom, and $E^{tot}(H_2)$ is the total energy of an isolated $H_2$ molecule in vacuum. First, we considered the case of nanovoids containing a single H adatom, and show the calculated trapping energies of H (i.e., $E(H_1, V_mH_1)$ with $m = 1{\sim}8$) in Fig. 1c. We found that those trapping energies can be nicely categorized by the Wigner-Seitz square where H resides after relaxation, falling into five distinct energy levels, in direct correspondence to the five different types of Wigner-Seitz squares. These energy levels are found to be insensitive to the nanovoid size, which confirms the previous assumption of the H energetics on a Wigner-Seitz square being only affected by the six metal sites enclosing the square. Fig. 1c shows the energetic preference of an H adatom on different Wigner-Seitz squares, which is in close accordance with the occupancy preference (e.g., see Fig. 1b) obtained from the *ab initio* MD simulations.

**Interaction between H adatoms.** After clarifying the distinct energy level of a single H adatom in nanovoids (i.e., $V_mH_1$), we further investigated the energetics of multiple H in nanovoids. In the follows, we first assume that H atoms stay in the form of adatoms for simplicity of analysis. The mutual interaction between H adatoms in nanovoids can be quantified by defining an H-H interaction energy

$$E^{int}(V_mH_n) = E^{tot}(V_mH_n) + (n-1)E^{tot}(V_m) - \sum_{k=1}^{n} E^{tot}(V_mH_1^{S_k}), \qquad (2)$$

where $S_1$, $S_2$, ... $S_n$ indicate the sites of the $n$ H adatoms in the $V_mH_n$ cluster, and $E^{tot}(V_mH_1^{S_k})$ represents the total energy of the reference metal matrix containing a $V_mH_1^{S_k}$ cluster with the H adatom located at $S_k$ site.

To unravel the complex H-H interaction, we started by examining nanovoids containing two H, i.e., $V_mH_2$. Fig. 2a shows the corresponding pairwise H-H interaction $E^{int}(V_mH_2)$, from which we see that the H-H interaction is generally repulsive and decays rapidly as the H-H separation distance $d$ increases. Intriguingly, we found the pairwise H-H interaction on a nanovoid surface can be well described by a $d^{-5}$ power law, similar to H on free metal surfaces[51], i.e.:

$$E^{int}(V_mH_2) = A_s d^{-5}, \qquad (3)$$

where $A_s = 3.19 \text{ eV}/\text{Å}^{-5}$ is a fitted constant. Consequent from this strong repulsion, it is expected that H adatoms would shift toward tetrahedral interstitial sites on Wigner-Seitz squares to increase H-H separation distances upon addition of H, consistent with Fig. 1a which shows high H probability at tetrahedral interstitial sites. Also, a Wigner-Seitz square would prefer to accommodate only one H adatom until all other Wigner-Seitz squares are occupied by H adatom(s), which is in line with the results presented in Fig. 1b

**Energetics of H adatoms in nanovoids.** For the general case of $V_mH_n$ (with $n \geq 2$), according to our definition of H-H interaction energy (cf. Eq. 2), the overall trapping energy of the $n$ number of H adatoms in $V_mH_n$ equals:

$$E(H_n, V_m H_n) = \sum_{k=1}^{n} E(H_1, V_m H_1^{S_k}) + E^{\text{int}}(V_m H_n). \quad (4)$$

If we assume that the H-H interaction, $E^{\text{int}}(V_m H_n)$, remains pairwise in nature and described by the same $d^{-5}$ law as in Eq. 3, we can then represent the overall trapping energy as:

$$E(H_n, V_m H_n) \cong \sum_{k=1}^{n} E_k^{ij} + A_s \sum_{k<l}^{n} d_{kl}^{-5}, \quad (5)$$

where $d_{kl}$ denotes the distance between two H adatoms respectively at $S_k$ and $S_l$ sites. As previously demonstrated (cf. Fig. 1c), $E(H_1, V_m H_1^{S_k})$ in Eq. 4 would conform to one of the five energy levels, being directly prescribed by the Wigner-Seitz square where site $S_k$ is located, denoted as $E_k^{ij}$.

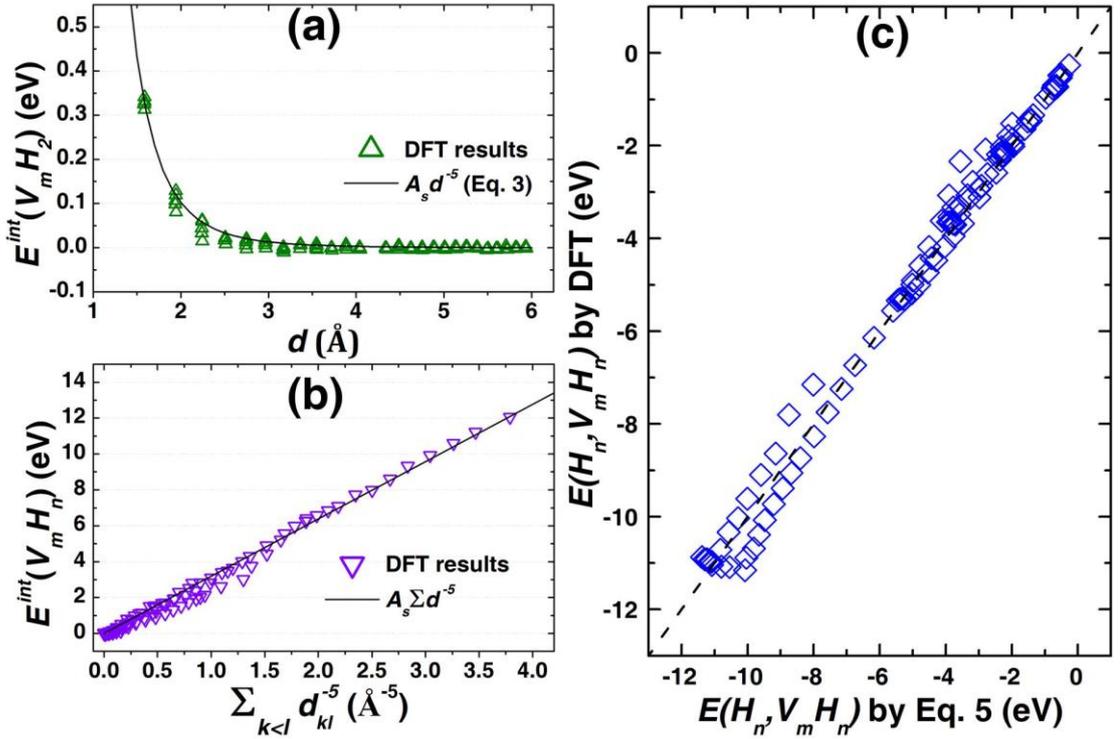

Figure 2 (a) Pairwise H-H interaction energies in different nanovoids, symbols are DFT data calculated using Eq. 2, solid lines are fitted using $d^{-5}$ power function, where $d$ is separation distance between two H adatoms at different tetrahedral interstitial sites, averaged among all symmetrically irreducible H-H pairs. (b) Interaction energies among multiple H adatoms in the most stable $V_m H_n$ clusters, as a function of the sum of pairwise $d^{-5}$. (c) Overall trapping energies from DFT calculations (symbols) against those predicted by Eq. 5.

To verify Eq. 5, we calculated trapping energies for $V_m H_n$ clusters[a]. We screened different candidate structures that selected randomly from the *ab initio* MD trajectories (15 structures for each $V_m H_n$), and that constructed by manually adding H on Wigner-Seitz squares with the lowest energy level, as well as that by adjusting H position to minimize Eq. 5. We found, in most cases, the minimizing Eq. 5 method can identify structures with the lowest energies. In Fig. 2b we also show that the interaction energy among multiple H adatoms can be well estimated by summing up pairwise $d^{-5}$ interactions, i.e., the adatom interaction term in Eq. 5. More importantly, as demonstrate in Fig. 2c, the overall trapping energies $E(H_n, V_mH_n)$ predicted by Eq. 5 are in close agreement with DFT data. These results clearly show that Eq. 5 captures the physical essence of multiple H adsorption on nanovoid surfaces, providing a simple but effective framework for determining stable structures of multiple H adatoms in nanovoids.

---

[a] Note that those clusters contain only H adatoms with no $H_2$ molecule formation, in order to just limit the focus on H adatoms.

Eq. 5 allow us to predict H trapping energies for any $V_mH_n$ cluster based on H structures. However, DFT relaxations are still indispensable for determining accurate H adatom sites and their mutual separation, which makes Eq. 5 less practical. To simplify the problem, yet without losing physical generality, two approximations are introduced here based on the above DFT results: i) H adatoms sequentially fill Wigner-Seitz squares with the lowest energy, and are uniformly distributed on the surface to maximize H-H distances; ii) each H adatom has six nearest H neighbors (i.e., close-packed distribution), and interactions between non-nearest adatoms are neglected considering the $d^{-5}$ rapid decay. In this way, $d = \left(\frac{2a}{\sqrt{3}n}\right)^{0.5}$ is the nearest H-H distance where $a$ is surface area of the nanovoid. Consequently, the multiple H-H interaction energy, i.e., the second term in the right side of Eq. 5, can be simplified to:

$$E^{int}(V_mH_n) \cong \frac{1}{2} A_s 6n \left(\frac{\sqrt{3}n}{2a}\right)^{2.5}. \qquad (6)$$

According to Eqs. 1, 5 and 6, the trapping energy of $n^{th}$ H adatom in a $V_mH_n$ cluster is given by:

$$E(H_1, V_mH_n) = E_k^{ij} + \frac{\partial E^{int}(V_mH_n)}{\partial n} \cong E_k^{ij} + 7.3 A_s \left(\frac{n}{a}\right)^{2.5}. \qquad (7)$$

Additional details related to the above can be found in Supplementary Section S2. Eq. 7 provides quantitative prediction of the trapping energies of H adatoms simply knowing the surface H density, $n/a$. The predictions from Eq. 7 are shown in Fig. 3 (blue lines) in comparison with the DFT-calculated trapping energies, showing very consistent agreement. This confirms the validity of the two approximations in describing the general behaviors of H adatoms in nanovoids. One particular observation from Fig. 3 is that the H trapping energy generally exhibits a combination of stepwise increments and gradual climbing. The stepwise growth is related to H occupying different types of Wigner-Seitz squares (Fig. 1), corresponds to distinct energy levels (i.e., different $E_n^{ij}$). Meanwhile the gradual climbing in the trapping energy results from the decreasing in the H-H distance and thus the increasing H-H repulsion. As H adatoms continue to populate the nanovoid, eventually the trapping energy will reach the energy of an interstitial H in bulk metal lattice, $E_{Bulk}^H = 0.92$ eV, when the nanovoid surface will be fully saturated, namely the surface H density researching its maximum. Under such condition, $E_n^{ij}$ will assume the highest energy level, i.e., $E_n^{ij=10}$ (see Fig. 1c), and the maximum density can be calculated (by $E(H_1, V_mH_n) = E_{Bulk}^H$) to be 0.304 H/Å$^{-2}$, corresponding to a nearest-neighboring H-H distance of 1.95 Å, in line with the H-H pairing state (1.94 Å, see Supplementary Section S1) observed in *ab initio* MD simulations.

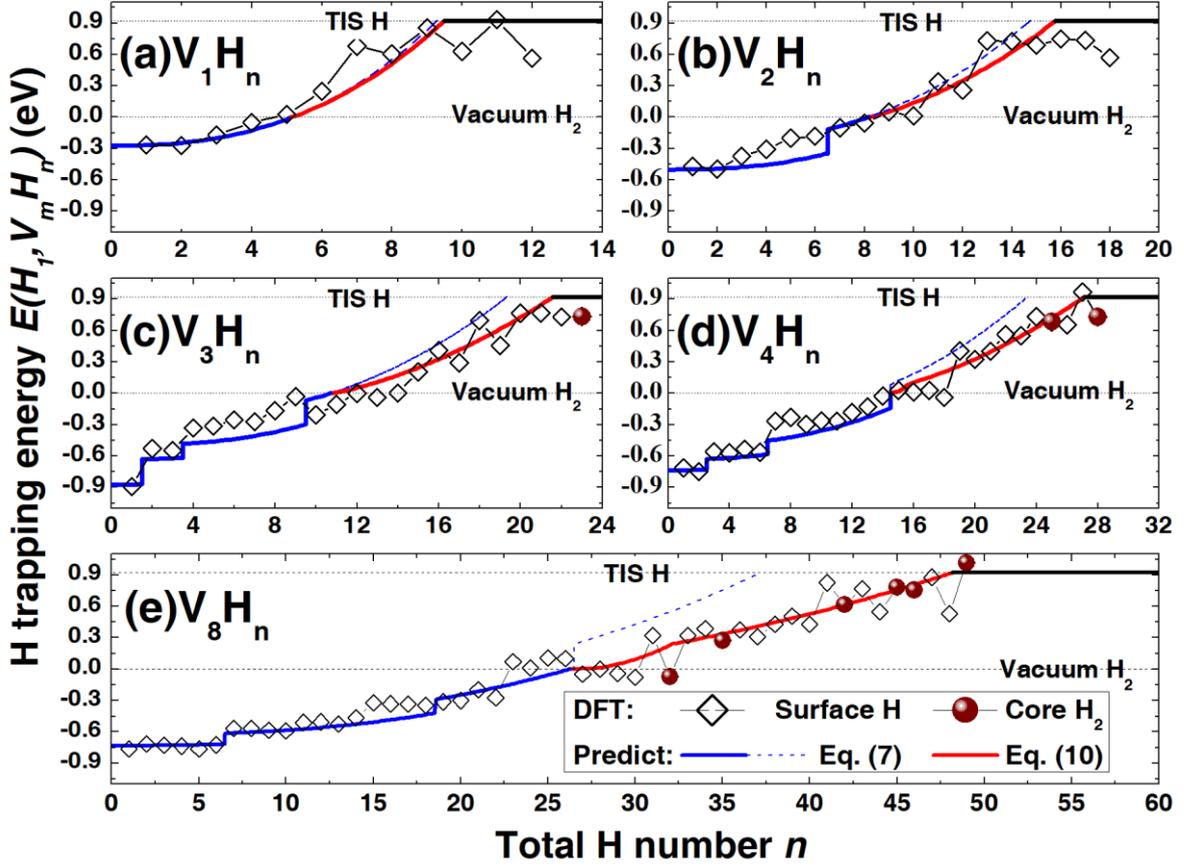

Figure 3 H trapping energy as a function of total H number in $V_1$-$V_4$ (a-d) and $V_8$ (e) nanovoids. DFT results are shown using open (for H adatoms) and solid (when forming a new $H_2$ molecule) symbols. Blue lines are model predictions given by Eq. 7, red lines are predictions given by Eq. 10. The lines marked "TIS H" and "Vaccum $H_2$" represent a H atom at the tetrahedral interstitial site in bulk metal lattice and in a $H_2$ molecule in vacuum, respectively.

**$H_2$ molecule formation in nanovoids.** In our above analysis, H atoms are assumed to stay in nanovoid in the form of adatoms. However, as seen in Fig. 3, with the continuous introduction of H into a nanovoid, the trapping energy of H adatom becomes positive, suggesting the possibility of $H_2$ molecule formation. Indeed, $H_2$ molecule formation has been observed in DFT calculations in large nanovoids (highlighted by solid symbols in Fig. 3). Further from Fig. 3, we note that along with the formation of H$_2$ molecules, notable deviation between the predictions (of H trapping energies) from Eq. 7 and the DFT data appears. Such deviation is well expected, as Eq. 7 is only applicable for describing H adatoms. To remedy such discrepancy, we need to account for H$_2$ molecule formation in our model. Noting that H$_2$ molecules in the nanovoid core can be characterized by the equation of states as (see details in Supplementary Section S2):

$$p = A_c \left(\frac{n_c}{v}\right)^3, \qquad (8)$$

where $p$ is pressure, $v$ is core volume of the nanovoid, $n_c$ is twice the number of $H_2$ molecules presented in the nanovoid, and $A_c = 8.01 \text{ eV/Å}^{-6}$ is a constant that well fitted to both experimental[52,53] and our DFT results. The trapping energies corresponding to molecular H in the nanovoid core can be expressed by:

$$E(H_1, V_m H_{n_c}) = \int_0^p \left(\frac{\partial v}{\partial n_c}\right)_P dP = \frac{3}{2} A_c \left(\frac{n_c}{v}\right)^2, \tag{9}$$

which depends on volumetric H density, $n_c/v$, in the nanovoid core. Denoting the number of H adatoms on the nanovoid surface as $n_s$, we have the total number of H in the nanovoid being $n = n_s + n_c$. Combining Eqs. 7 and 9, we can determine the partition of H in the states of adatom and molecule:

$$E(H_1, V_m H_n) = \frac{3}{2} A_c \left(\frac{n_c}{v}\right)^2 = E_{n_s}^{ij} + 7.3 A_s \left(\frac{n_s}{a}\right)^{2.5}. \tag{10}$$

From the above, the further evolution of H trapping energy in the event of H$_2$ molecule formation can then be obtained, which is illustrated as the red curves in Fig. 3. We see this new model prediction yields excellent agreement with the DFT data for the regime of H trapping energy being positive. Moreover, the above framework can be readily extended to H bubbling in a nanovoid at finite temperatures and varying chemical environments, by equaling the chemical potential of H in nanovoid core, $\mu_{core}^H$, with that of bulk H, $\mu_B^H$:

$$\mu_{core}^H = \int_0^p \left(\frac{\partial v}{\partial n_c}\right)_P dP = \mu_B^H = E_{Bulk}^H + k_B T \ln\left(\frac{C_H}{1-C_H}\right) \tag{11}$$

where $k_B$ is Boltzmann constant, $T$ denotes the temperature, $C_H$ is bulk H concentration, and $E_{Bulk}^H$ (= 0.92 eV) is the trapping energy of H interstitial in bulk W lattice. Note the equation of states for high pressure $H_2$ is relatively insensitive to temperatures[54]. Consequently, combining Eqs. 8 and 11, the H bubbling pressure under thermodynamic equilibrium can be obtained as:

$$p = \frac{1}{\sqrt{A_c}} \left[\frac{2}{3}\left(E_{Bulk}^H + k_B T \ln\left(\frac{C_H}{1-C_H}\right)\right)\right]^{\frac{3}{2}}, \tag{12}$$

which solely depends on temperature, the concentration and energy state of H in the bulk lattice. In the presence of high H concentration and low temperatures, the bubble pressure can be high enough to induce spontaneous bubble growth via mechanisms like loop-punching (see Supplementary Section S4).

**Discussion**

The excellent agreement between model prediction and DFT data (cf. Figs. 2-3) evidences that our model captures the fundamental physics underlying H trapping and interaction in nanovoids. It is important to note that the model parameters $E_n^{ij}$, $A_s$ and $A_c$ are not sensitive to the size or configuration of nanovoid, and thus the model is readily applicable to larger systems to examine H bubble formation. Furthermore, preliminary calculations have been performed for Mo, Cr and $\alpha$-Fe systems with similar H behaviors demonstrated (see Fig. S8 in Supplementary Section S3), confirming the generality of our model for application in other BCC metals.

The proposed model accurately predicts H trapping configurations, H energetics and H$_2$ molecule formation in nanovoids, with a finite set of DFT calculations. Such predictions may serve as critical benchmarks for developing new metal-H empirical interatomic potentials for classical MD simulations (see Supplementary Section S4). Meanwhile they provide atomic-precision data to feed into coarse-grained methods such as kinetic Monte Carlo simulation[55] or mean-field approaches[24], enabling a multiscale approach that directly bridges atomistic data with macroscopic experiments, such as thermal desorption spectroscopy (TDS) results. Below we presented one example comparing our predictive model based multiscale simulations with recent deuterium (*D*) TDS experiments conducted by Zibrov et al. and Ryabtsev et al[19, 20]. In these two experiments, irradiation damaged W samples were treated by particular annealing procedures at 550K and 800K respectively. With vacancies being mobile only beyond 550K[56, 57], it is postulated[19, 20] that these two annealing temperatures would render irradiation induced defects in W samples as small and large nanovoids respectively. These annealed samples were subsequently implanted by low-energy *D* ions, followed by TDS measurements with different heating rates. The TDS spectra data corresponding to radiation defects (reproduced from Refs [19, 20]) were shown in Fig. 4.

Combining our predictive model, primary irradiation damage simulations, and object kinetic Monte Carlo method, we carried out multiscale simulations to reproduce the experimental TDS results (details please see Methods and Supplementary Section S4). To make a direct comparison, same irradiation and annealing conditions as those in the experimental studies[19, 20] were used in our simulations. As seen in Fig. 4, the simulated TDS curves (lines) well match the experimental results (symbols) for the temperature range in which H-nanovoid interaction dominates. In particular, the simulated desorption peaks in Fig. 4a correspond to *D* release from small nanovoids (mostly $V_1 - V_2$), while the desorption peaks in Fig. 4b are attributed to large nanovoids (mostly $V_6 - V_{15}$), confirming the experimental speculations[19, 20]. Aside from some minor discrepancy at low temperatures which may contributed by other defects like dislocations or grain boundaries[19, 20], our simulations accurately reproduce the experimental observations.

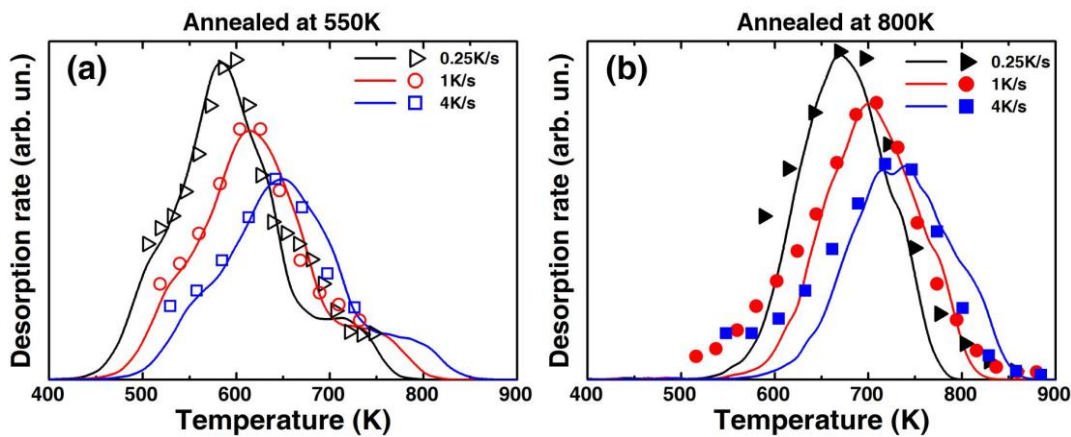

Figure 4. Thermal desorption spectra of deuterium from W after irradiated by 10 keV/D ions to the fluence of $3 \times 10^{19}$ D/m$^2$, annealed at (a) 550 K and (b) 800K for 5 min to render small and large nanovoids, implanted with 0.67 keV/D ions to the fluence of $10^{19}$ D/m$^2$, then annealed with different heating rates. Lines are multiscale modelling results, while symbols are experimental data from Zibrov et al. (open) and Ryabtsev et al. (solid) [19, 20].

In summary, the present study, for the first time, explicitly demonstrated sequential adsorption of hydrogen adatoms on Wigner-Seitz squares of nanovoids with distinct energy levels. It clarifies fundamental physical rules governing H trapping, interaction and bubbling in nanovoids in BCC metals. A comprehensive modeling framework was established to enable accurate predictions of H energetics and H$_2$ molecule formation in nanovoids. Our study offers long-sought mechanistic insights crucial for understanding hydrogen-induced damages in structural materials, and provides essential predictive tools for developing new H-metal interatomic potentials, and multiscale modeling of H bubble nucleation and growth.

**Methods**

**First-principles density functional theory (DFT) calculations.** First-principles DFT calculations were performed employing the Vienna *ab initio* simulation package (VASP) [58, 59] with Blochl's projector augmented wave (PAW) potential method[60]. All the 5d and 6s electrons of metal and 1s electron of H were treated as valence electrons. The exchange-correlation energy functional was described with the generalized gradient approximation (GGA) as parameterized by Perdew-Wang (PW91)[61, 62]. A super-cell containing 128 lattice points (a $4 \times 4 \times 4$ duplicate of a conventional BCC unit cell) was used in our calculations.

Nanovoids were constructed with stable structures from previous DFT study[63]. Relaxation of atomic positions and super-cell shapes and sizes was performed for all calculations, except for those examining pairwise H-H interaction, where all atoms were fixed to avoid H-H separation adjustment from relaxation. The convergence criteria for energy and atomic force were set as $10^{-6}$ eV and 0.01 eV/Å respectively. A 500 eV plane wave cutoff and a $3 \times 3 \times 3$ *k*-point grid obtained using the Monkhorst-Pack method[58] were used. Benchmark calculations with increased super-cell size, cutoff energy and *k*-point density, as well as zero-point energy correction for H, have been carried out, and negligible influence on our results was found.

***Ab initio* molecular dynamics (MD) simulations.** *Ab initio* MD simulations were performed in the canonical (NVT) ensemble with Nose-Hoover thermostat using the VASP code. A lower cutoff energy (350 eV) and a $1 \times 1 \times 1$ *k*-point grid were adopted. The Verlet algorithm is used for integration of newton's equations of motion. All systems were simulated at the temperature of 600K with a time step of 1 femtosecond. H atoms were randomly added into nanovoids at a rate of 1 atom per 5 picoseconds. To avoid potential influence of initial H addition positions, results for the first 2 picoseconds after every H addition were excluded from the spatial distribution analysis.

**Multiscale simulations.** Thermal desorption spectra of H isotopes were simulated by a quantitative multiscale modelling approach, which incorporates atomistic scale H-nanovoid interactions, irradiation induced primary damages, and large-scale object kinetic Monte Carlo (OKMC) simulations. General algorithms of the OKMC method and parameterizations of defects were described in detail elsewhere[39, 55, 64, 65]. A $60 \times 60 \times 500\ nm^3$ box was used in all OKMC simulations, with periodic boundary conditions applied on the first two dimensions and H allowed to desorb at the surface of the third dimension. The interactions between H and a nanovoids were parameterized using the predictive model conveyed by Eq. 10. Primary irradiation damage databases were tabulated using the binary collision Monte Carlo code IM3D[66], and invoked during OKMC simulations of *D*. The kinetic energies of *D*, irradiation fluxes, irradiation time, and temperatures were calibrated according to the corresponding experimental conditions[19].

# Supplementary Information for Predictive Model of Hydrogen Trapping and Bubbling in Nanovoids in BCC Metals

Jie Hou[1,2,3], Xiang-Shan Kong[1], Xuebang Wu[1], Jun Song[3], C. S. Liu[1]

[1]Key Laboratory of Materials Physics, Institute of Solid State Physics, Chinese Academy of Sciences, P. O. Box 1129, Hefei 23031, P. R. China

[2]University of Science and Technology of China, Hefei 230026, P. R. China

[3]Department of Mining and Materials Engineering, McGill University, Montreal, Quebec H3A 0C5, Canada.

## S1. *Ab initio* Molecular Dynamics

### S1.1. H-H Radial distribution

H-H radial distribution density function is used to characterize the pairing state of H atoms in nanovoids from *ab initio* Molecular Dynamics (MD) simulations, with the radial distribution density:

$$g(r) = \frac{1}{4\pi r^2 \rho_H} \frac{dn_H}{dr}, \quad (S1)$$

where $\rho_H$ is volumetric density of H, $n_H$ is the mean number of H atoms in a shell of width $dr$ at distance $r$ from a center H atom. Figure S1 shows H-H radial distributions in different nanovoids. Two distinct H-H pairing states can be identified, one with an average peak position located around 0.75 Å that is observed for large nanovoids (with $V_3$ having a transient state at 0.89 Å), corresponding to the formation of $H_2$ molecules (i.e., $H_2$ bond length = 0.74 Å), while the other peaks at 1.94 Å, related to H adatom adsorption on the inner nanovoid surface (observed in all nanovoids).

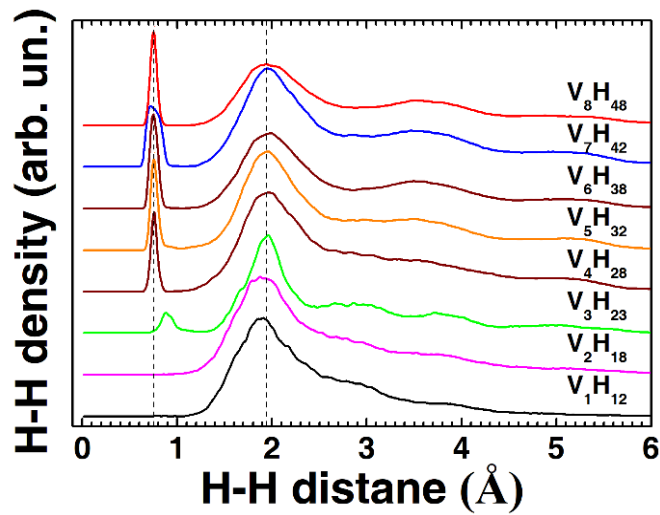

**Figure S1**. H-H radial distribution density in nanovoids at a sample temperature of 600K. Curves are vertically shifted and the vertical axis is in arbitrary unit (arb. un.).

### S1.2. Spatial distribution of H

The spatial distribution of H in a nanovoid is analyzed on the basis of the H probability density function, i.e., the probability of finding an H atom in unit volume around different locations. Figure S2 shows the H probability density function in different nanovoids, obtained from *ab-initio* MD simulations at 600K, clearly indicating H adatoms absorbing solely on the Wigner-Seitz squares of nanovoid surfaces, preferentially around vertices (tetrahedral interstitial sites) of the squares. For large nanovoids, e.g., $V_4$ and $V_8$, H distribution in the core region becomes visible, attributing to $H_2$ molecule formation.

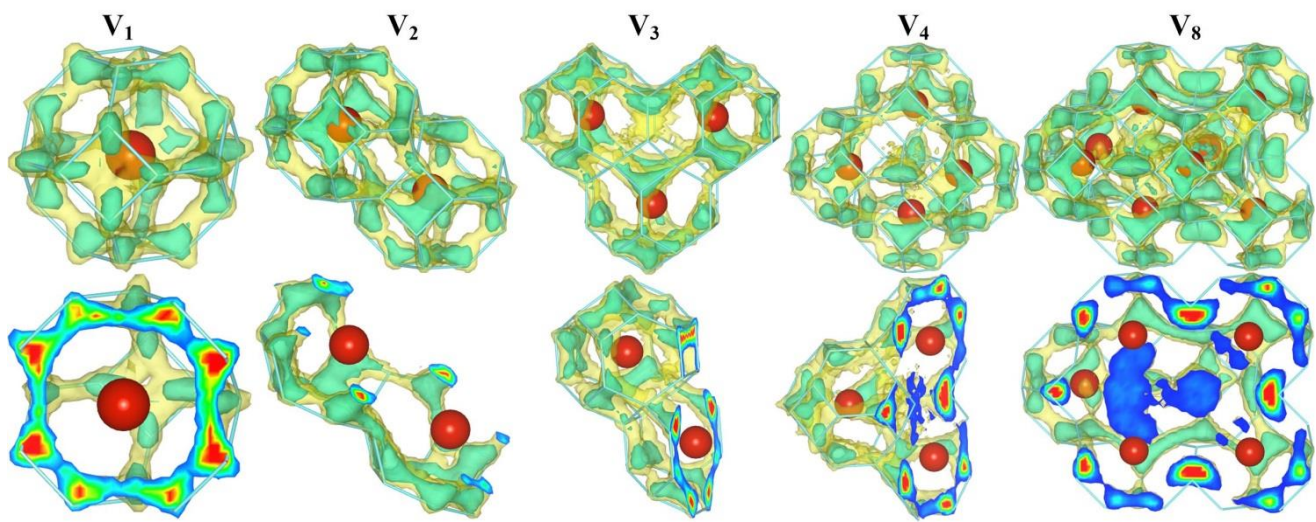

**Figure S2**. H probability density in different nanovoids during H adding (i.e., summarized results for $V_m H_n$ with different $n$), with the upper row showing the two isosurfaces of the density, while the lower row showing the corresponding cross-sections. The red spheres indicate the cores of those vacancies constituting the nanovoid, while bright green lines indicate edges of Wigner-Seitz cells on the nanovoid surface. Tungsten (W) atoms are fixed at ideal BCC sites in these *ab-initio* MD simulations.

As H atoms were introduced into the nanovoid, the average number of H atoms on each type (*ij*) of Wigner-Seitz squares was monitored. As demonstrated in Figure S3, the preference of Wigner-Seitz squares follows the sequence of *ij* = 22, 21, 12, 11 and 10. It is also observed that each Wigner-Seitz square prefers to accommodate only one H adatom unless all Wigner-Seitz squares are occupied by H adatoms. As H continues to populate the nanovoid, multiple H adatoms are expected in Wigner-Seitz squares and eventually the crowdedness will drive H adatoms to combine into $H_2$ molecules in the nanovoid core.

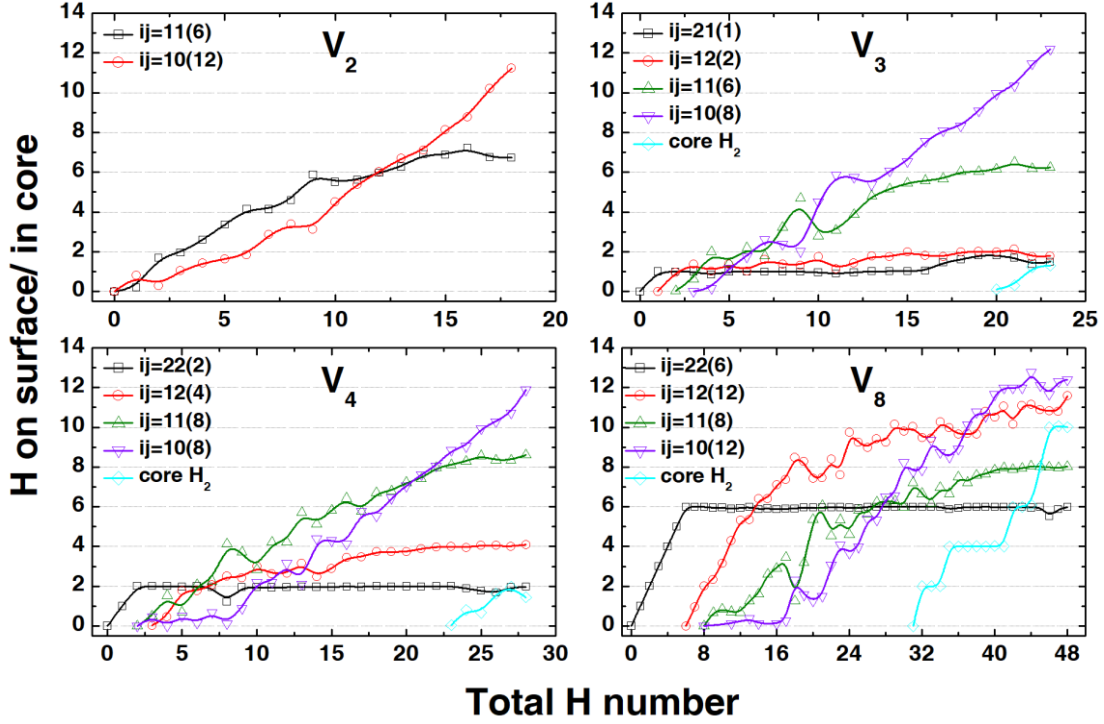

**Figure S3**. The evolution of average numbers of H on different types (*ij*) of Wigner-Seitz squares (as adatoms) or in core (as molecules) in different nanovoids at 600K during continuous H introduction in *ab-initio* MD simulations, shown in parenthesis are corresponding Wigner-Seitz square numbers, results for $V_1$ are not shown because $V_1$ only has ij=10 type Wigner-Seitz squares and no $H_2$ molecule formation.

**Table SI**: Numbers of Wigner-Seitz squares on surfaces of $V_1$-$V_8$ nanovoids in BCC metals.

|  | *ij*=10 | 11 | 12 | 21 | 22 | Sum |
|---|---|---|---|---|---|---|
| $V_1$ | 6 | | | | | 6 |
| $V_2$ | 6 | 6 | | | | 12 |
| $V_3$ | 8 | 6 | 2 | 1 | | 17 |
| $V_4$ | 8 | 8 | 4 | | 2 | 22 |
| $V_5$ | 8 | 10 | 6 | | 2 | 26 |
| $V_6$ | 10 | 8 | 8 | | 4 | 30 |
| $V_7$ | 10 | 10 | 10 | | 4 | 34 |
| $V_8$ | 12 | 8 | 12 | | 6 | 38 |

## S2. Further details on H adsorption and H energetics prediction in nanovoids

### S2.1. Adsorption sequence of H adatoms at 0K

In the results shown in Figure S3, we can see clear influence of thermal energy. Therefore we further examined H adsorption in nanovoids at 0K to avoid thermal perturbation. Figure S4 shows the number of H on different Wigner-Seitz squares during continuous H addition at 0K. Here we only consider the most stable structure for each $V_mH_n$ cluster (a structural demonstration of nanovoids filled up with H is shown in Figure S5). The results in Figure S4 further (and more strongly) evidence the preference of H adatoms on different Wigner-Seitz squares, namely H adatoms sequentially filling the Wigner-Seitz squares accordingly to the energy levels (from low to high). There remain some exceptions, which are attributed to the nonuniform distribution of Wigner-Seitz squares. Further it is worthy to note that most of those exceptions occurred for the adsorption on $ij$=21 and $ij$=12 Wigner-Seitz squares, whose energy levels for H adsorption are very close. Consequently, those exceptions do not influence our model prediction of H energetics.

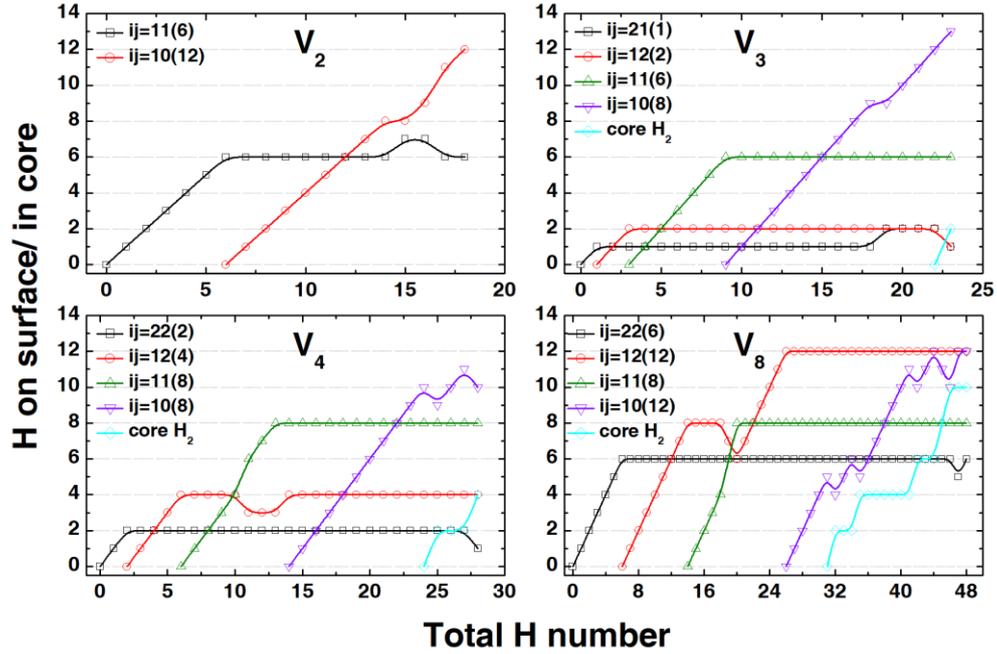

**Figure S4**. The numbers of H on different types ($ij$) of Wigner-Seitz squares (as adatoms) or in core (as molecules) in different nanovoids at 0K, shown in parenthesis are corresponding Wigner-Seitz square numbers

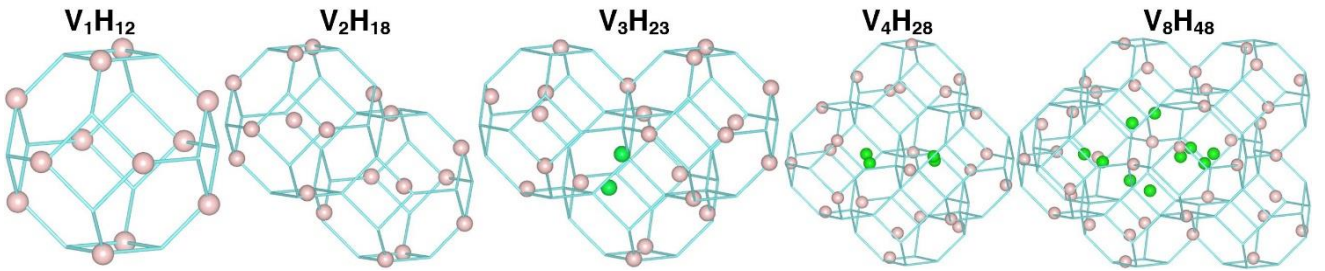

**Figure S5**. The most stable configurations $V_1$-$V_4$ and $V_8$ nanovoids filled up with H. Bright green lines are edges of Wigner-Seitz cells of vacancies, pink spheres are H adatoms, green spheres are $H_2$ molecules.

## S2.2. Surface distribution of H adatoms

As discussed in the main text, the total interaction energy among multiple H adatoms in a $V_mH_n$ cluster, $E^{int}(V_mH_n)$, can be well estimated by the summation of all pairwise H-H interaction energies. With a close-packed surface H distribution assumed and interactions between non-nearest adatoms neglected, $E^{int}(V_m H_n)$ becomes:

$$E^{int}(V_m H_n) = \frac{1}{2} A_s \sum_{l=1}^{n} \sum_{k=1}^{n} d_{kl}^{-5} = \frac{1}{2} A_s 6n \left(\frac{\sqrt{3n}}{2a}\right)^{2.5}. \tag{S2}$$

In this way, the addition of the $n^{th}$ H adatom in a $V_m H_n$ cluster will increase the H-H interaction energy by:

$$E^{int}(V_m H_n) - E^{int}(V_m H_{n-1}) \approx \frac{\partial E^{int}(V_m H_n)}{\partial n} \approx 7.3 A_s \left(\frac{n}{a}\right)^{2.5}, \tag{S3}$$

Note that, the stable surface H-H distribution is not always close-packed. For instance, in the case that actual distribution is close to a simple square distribution, where each H adatom has 4 nearest neighbors. The above equations become:

$$E^{int}(V_m H_n) = \frac{1}{2} A_s 4n \left(\frac{n}{a}\right)^{2.5}, \tag{S4}$$

$$E^{int}(V_m H_n) - E^{int}(V_m H_{n-1}) \approx 7 A_s \left(\frac{n}{a}\right)^{2.5}, \tag{S5}$$

which is very close to the results derived from close-packed distribution. This is because reducing the nearest neighbors from 6 to 4 will in turn reduce the nearest H-H distance from $\left(\frac{2a}{\sqrt{3n}}\right)^{0.5}$ to $\left(\frac{a}{n}\right)^{0.5}$. These two effects basically even each other out, consequently rendering the actual surface distribution of H adatoms not important in affecting the energetics.

### S2.3. Equation of states of H₂ molecules in nanovoids

The equation of state (EOS) of H₂, which describes the pressure-density relationship of pressurized H₂, was determined by DFT calculations (up to 120 GPa). According to previous theoretical studies[1], the Pa3 structure is the most stable arrangement for H₂ under 150 GPa, and the structure dependence of the EOS is rather weak[1]. Therefore, we only considered the Pa3 structure in our DFT calculations. The calculated EOS is shown in Figure S6 along with available experimental results reported in previous studies[2, 3], demonstrating good agreement between our DFT results and the experimental data.

As suggested by many previous studies[3-5], the EOS can be estimated by a cubic function of H density, in the general form:

$$p = A_c \rho^3 + O(\rho), \tag{S6}$$

where $\rho$ is H density, and $O(\rho)$ are low-order terms of the H density, usually trivial at GPa scale. Therefore, we just consider the cubic term in the present work for simplicity. As shown in Figure S6, the cubic function with $A_c = 8.01\ eV/\text{Å}^{-6}$ provides an excellent fitting to both experimental results and our DFT data of EOS.

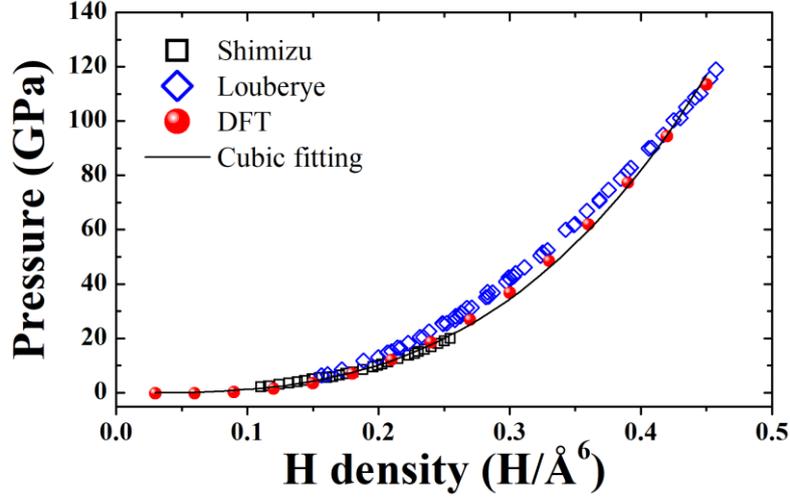

**Figure S6**. Equation of state (EOS) of $H_2$, with open symbols being previous experimental studies by Louberye et al.[2] and by Shimizu et al.[3], solid spheres being our DFT results, and the black line representing cubic fitting of our DFT results.

### *S2.4. Geometric parameters of nanovoids*

Spherical shape is usually the most stable shape for nanovoids [6]. Therefore, we assume all nanovoids are spheres with same partial volume to bulk W to simplify our calculations. A schematic of surface and core regions of a spherical nanovoid is illustrated in Figure S7. The radius of a $V_m$ nanovoid is given by $= a_0 \left(\frac{3m}{8\pi}\right)^{\frac{1}{3}}$, where $a_0 = 3.165$ Å is the W lattice constant, and the surface area of the nanovoid simply equals the area of the sphere. Note that, according to Figure S1, H adatoms on nanovoid surfaces have a minimum interaction distance of $d_H = 1.94$ Å. As a result, the nanovoid core is therefore defined as a sphere with a radius of $r - \frac{1}{2}d_H$. Table SII shows the surface area and core volume calculated following the above approach. Substituting these two parameters into the model we proposed in the main text (cf. Eq. 10), the maximum H capacities (number of H when the H trapping energy equals that at tetrahedral interstitial site in bulk, i.e., 0.92 eV) on the surface and in the core region can be predicted, also enlisted in Table SII, and compared to the maximum H capacities determined from DFT calculations.

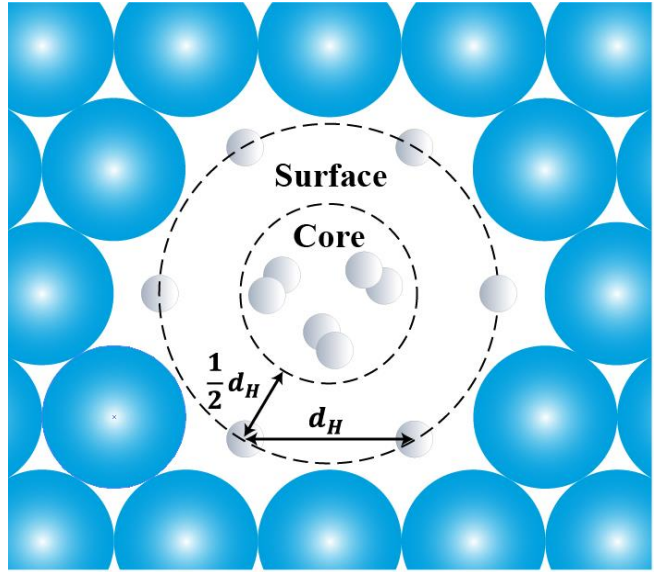

**Figure S7**. Schematic of surface and core regions of a nanovoid filled with H. $d_H = 1.94$ Å is the minimum distance between H adatoms according to **Figure S1**.

**Table SII**. Surface area ($a$), core volume ($v$) of spherical nanovoids, along with comparison between predicted and DFT calculated H capacities for the two regions.

|  | $a(\text{Å}^2)$ | $v(\text{Å}^3)$ | Predict capacity | | DFT capacity | |
| --- | --- | --- | --- | --- | --- | --- |
|  |  |  | Surface | Core | Surface | Core |
| $V_1$ | 30.5 | 0.7 | 9.3 | 0.2 | 12 | 0 |
| $V_2$ | 48.4 | 3.7 | 14.7 | 1.1 | 18 | 0 |
| $V_3$ | 63.5 | 8.1 | 19.3 | 2.4 | 21 | 2 |
| $V_4$ | 76.9 | 14.0 | 23.4 | 3.9 | 24 | 4 |
| $V_8$ | 122.1 | 39.7 | 37.2 | 11.5 | 38 | 10 |

## S3. Generality of the Proposed Model to Other BCC Metal Systems

Preliminary studies on other representative BCC metal systems, including Mo, Cr, and α-Fe (VB group BCC metals were not considered because they usually form metal-hydrides rather than H bubbles), were performed to demonstrate the general applicability of our proposed model. Similar to what we did for the W system, the energy states of single H adatoms and their mutual interactions at different separation distances were examined. The results for Mo, Cr, and α-Fe are shown in Figure S8, along with comparison to those for W. We can see that the energetic behaviors for H adatoms in nanovoids are similar in different BCC metals, i.e., individual adatoms occupy Wigner-Seitz squares with distinct energy levels and the pairwise H-H interaction energy is well approximated by a $d^{-5}$ power function of the separation distance $d$. Considering these similar behaviors of H adatoms in nanovoids, and that $H_2$ in nanovoid core is not influenced by the metal matrix, the model proposed in our study is expected to be generic and applicable to different BCC metals.

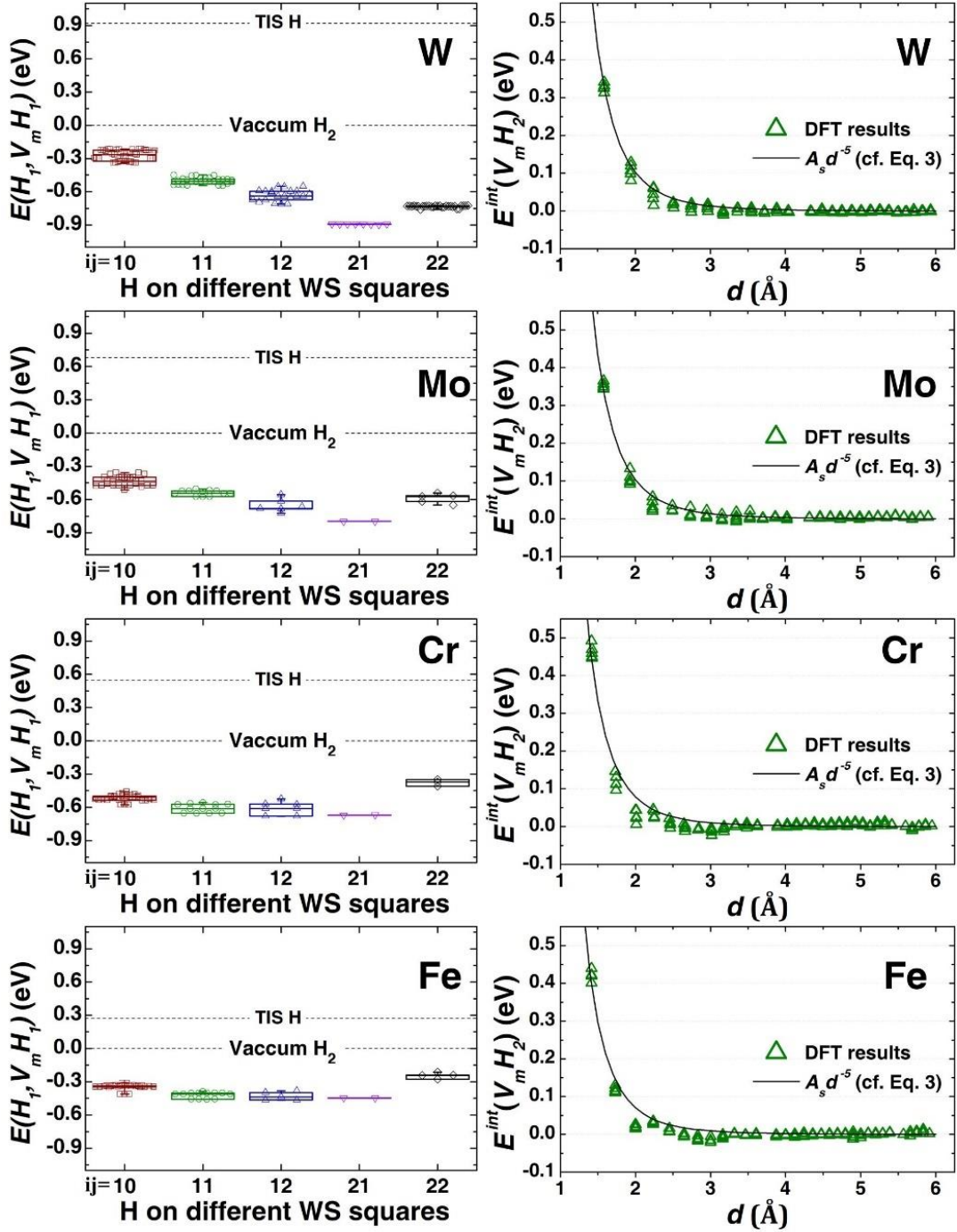

**Figure S8**. (Left panel) The energy states of a single H adatom on different Wigner-Seitz squares and (right panel) average pairwise interaction energy between two H adatoms at different separation distances in of $V_1$-$V_4$ and $V_8$ nanovoids in W, Mo Cr and Fe BCC metals. Solid lines are fitted curves using the $d^{-5}$ power function with $A_s = 3.19, 3.32, 2.52, 2.25$ eV/Å$^{-5}$, respectively.

**S4. From Proposed Predictive Model to Large-scale Modeling**

*S4.1. Details for multiscale modeling*

*Ab initio* parameters, including binding energies between H and self-interstitial atoms (SIA) or between interstitial H atoms, and migration energies for interstitial H atoms, vacancies, and SIA, were adopted from previous *ab-initio* studies[7-9]. Binding energies between H and nanovoids were obtained using the predictive model proposed in the present study. We first constructed different $V_m$ nanovoid configurations ($m < 200$) by removing W atoms within a sphere with different radii, then determined number of Wigner-Seitz squares on surfaces of these nanovoids. After that, binding energies between H and these nanovoids can be readily obtained by solving Eq. 10 in the main text. Energetic deuteron irradiation database, including depth distribution of stopped deuterium and three dimensional spatial correlation between deuterium and irradiation damages, were simulated and tabulated using the IM3D binary collision Monte Carlo code[10].

Figure S9 shows simulated deuterium thermal desorption rates contributed by different nanovoids, corresponding to the 1K/s data in cf. Figure 4 in the main text. The 550 K annealing procedure cannot cause significant vacancy clustering, leading to a deuterium desorption dominated by small nanovoids (monovacancies and divacancies). While after annealed at 800 K, most monovacancies and divacancies are aggregated into vacancy clusters. Therefore, the deuterium desorption is dominated by large nanovoids (mostly $V_6 - V_{15}$). Since deuterium trapping in small nanovoids are weaker compared with large nanovoids, the desorption peak temperatures related to small nanovoids are lower than that of large nanovoids.

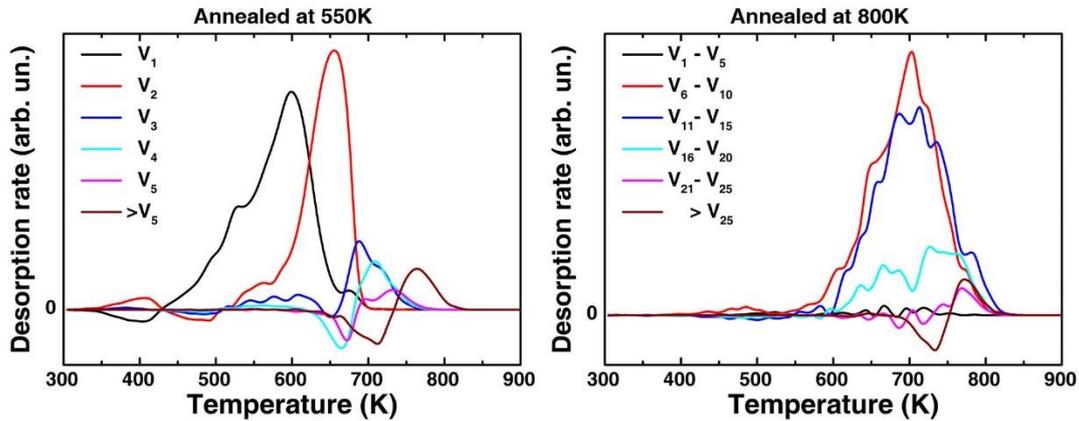

**Figure S9**. Simulated deuterium thermal desorption rates from different nanovoids in W after irradiated by 10 keV/D ions to the fluence of $3 \times 10^{19}$ D/m², annealed at (left) 550 K and (right) 800K for 5 min to render small and large nanovoids, implanted with 0.67 keV/D ions to the fluence of $10^{19}$ D/m², then annealed with 1 K/s heating rate.

*S4.2. Benchmarks for developing empirical potentials*

The availability and accuracy of empirical inter-atomic potentials are crucial for large-scale modelling of H-nanovoid interplay using classical molecular dynamics (MD) simulations. Brute-force DFT results of H-nanovoid binding energies, are common benchmarks in developing new metal-H inter-atomic potentials. However, it is computational expensive to obtain a comprehensive set of such data. The predictive model established in this work provides a quick route to obtain accurate estimation of H-nanovoid binding energies, thus accelerating the development of new metal-H interatomic potentials. Below we presented an example in Figure S10, illustrating the usage of the model to benchmark available interatomic potentials. As seen in Figure S10, the potential recently developed by Wang et al.[11] provides the best description of H-nanovoid interaction, but still generally underestimates H-nanovoid binding energies.

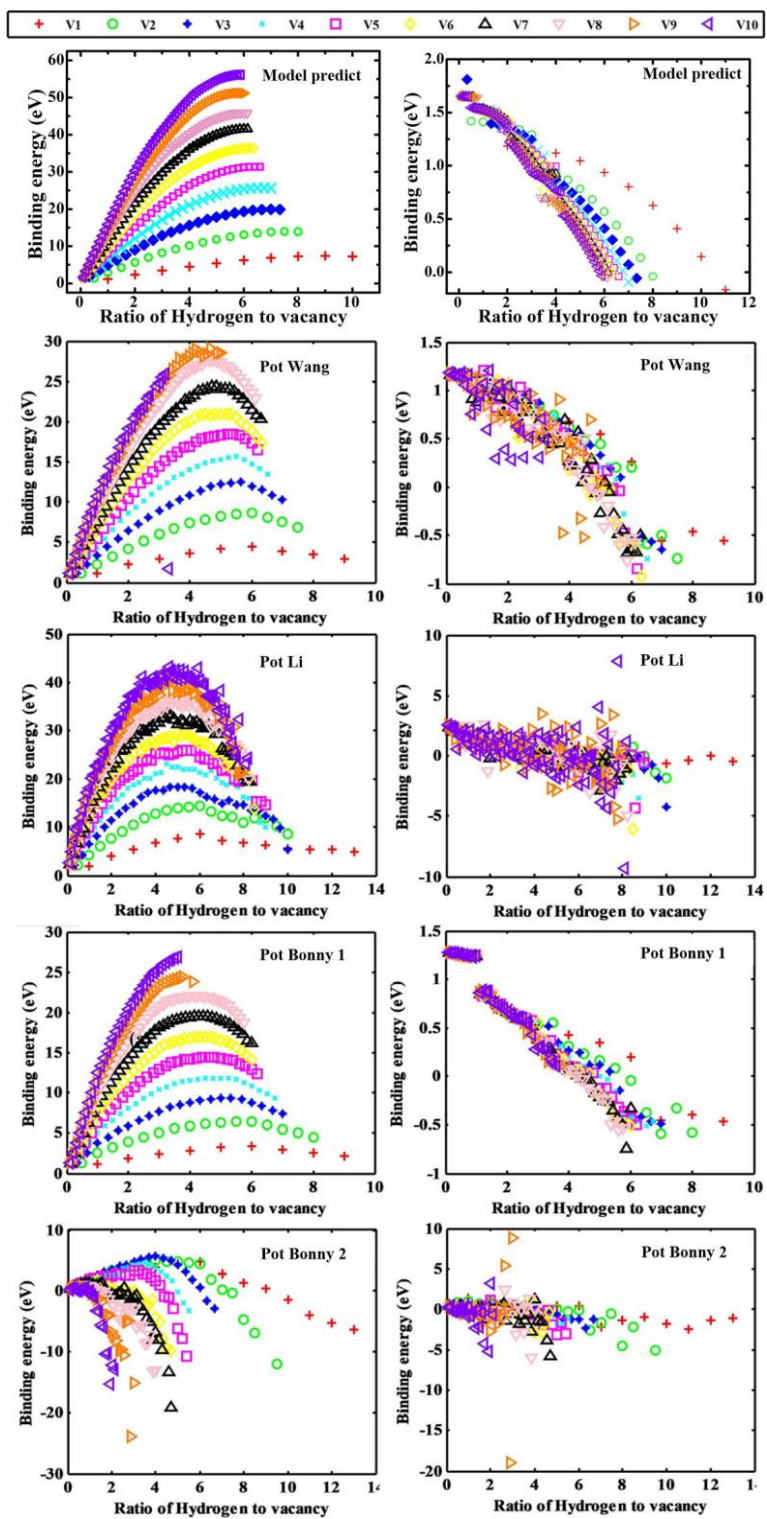

**Figure S10**. Binding energies between H and different nanovoids (i.e., energy difference between H in bulk and H in nanovoids) estimated by our model prediction and by different inter-atomic potentials developed by Wang et al[11], Li et al[12], and Bonny et al[13] (data adopted from Wang et al's work[11]). Left panels are the overall binding energies of all H in the corresponding nanovoid. Right panels are the binding energies of each H sequentially added in the nanovoid.

*S4.3 Spontaneous loop-punching under different temperature and H concentration*

Figure. S11a shows the bubble pressure under different conditions predicted by cf. Eq. 12 in the main text. The pressure increases rapidly with bulk H concentration, while decreases with temperature. According to Greenwood et al.[14], when the pressure inside a bubble is higher than a critical value, the nanobubble will spontaneously punch out dislocation loops to relieve the pressure. The loop-punching criterion is given by:

$$p > \frac{2\gamma}{R} + \frac{bG}{R}, \quad (S7)$$

where $\gamma$ is surface energy of metal, being $4\,J/m^2$ for close-packed {110} W surfaces[15], $b = \sqrt{3}a_0/2$ is the norm of the Burgers vector, $G = 160\,GPa$ is the shear modulus of W, and $R$ is the radius of the bubble. Once the bubble radius grows larger than a critical value that given by Eq. S7, which is shown in Figure S11b, the loop-punching process will become spontaneous and enables a rapidly growth of the nanobubble.

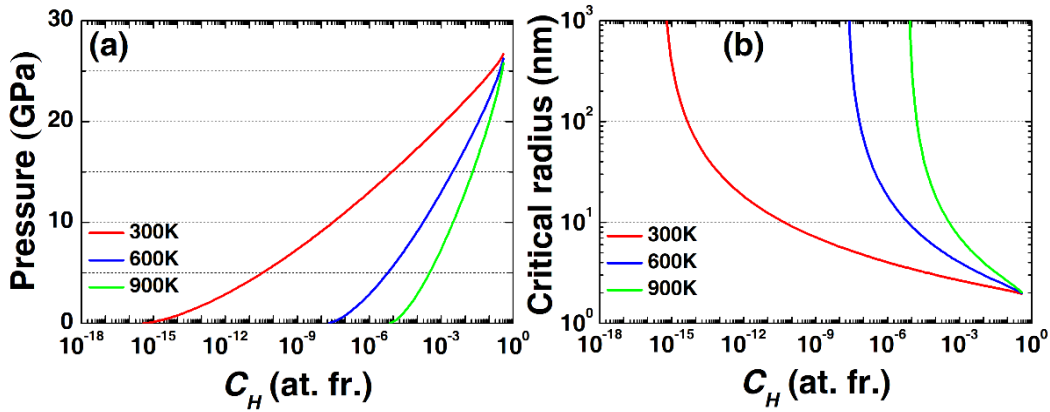

**Figure S11**. (a) equilibrium nanobubble pressure under different temperature and bulk H concentration. (b) critical nanobubble radius for spontaneous loop-punching under different temperature and bulk H concentration.